\documentclass[11pt,twoside]{article}

\usepackage{asp2006}
\usepackage{epsf}
\usepackage{psfig}
\usepackage{lscape}
\usepackage{graphicx}

\begin{document}
\title{Physical information derived from the internal structure in jets}  
\author{M. Perucho$^{1}$, A. P. Lobanov$^{1}$ and Y. Y. Kovalev$^{1,2}$}   
\affil{$^1$Max-Planck-Institut f\"ur Radioastronomie, Auf dem H\"ugel 69, 53121, Bonn, Germany}    
\affil{$^2$Astro Space Center of Lebedev Physical Institute, Profsoyuznaya 84/32, 117997 Moscow, Russia}

\begin{abstract}
We present the first results on the analysis of the structures
observed in the jet of the quasar 0836+710. We obtain the ridge
lines of the jet at different epochs and several frequencies. We
interpret the oscillatory structures obtained as waves that can be
attached to the growth of instabilities. We explain how to derive 
information on the nature and origin of these structures by
fitting together the ridge lines at different epochs and
frequencies. Finally we show the predictive power of this
approach: by generating an artificial wave and applying the
corresponding relativistic and projection effects we show that
apparent changes in the jet direction in the inner regions of jets
can be attached to the transversal motion of structures.
\end{abstract}

\section{Introduction}
Understanding the nature of the structures and motions observed in
jets is a crucial step in our knowledge of the physics of these
objects. On parsec scales, this has become feasible only recently,
using space VLBI (Very Long Baseline Interferometry) observations
with the VSOP (VLBI Space Observatory Programme) \citep{lz01}.
These observations revealed the presence of a double helical
structure inside the jet of 3C~273, which can be attributed to a
combination of the helical and elliptic modes of Kelvin-Helmholtz
(KH) instability. Numerical simulations further support this
interpretation \citep{pe06}.

The luminous quasar S5\,0836+710 at a redshift $z=2.16$ hosts a
powerful radio jet extending up to kiloparsec scales \citep{hu92}.
VLBI monitoring of the source \citep{ot98} has yielded estimates
of the bulk Lorentz factor $\gamma_\mathrm{j}=12$ and the viewing
angle $\theta_\mathrm{j}=3^\circ$ of the flow. The presence of an
instability developing in the jet is suggested by the kink
structures observed on milliarcsecond scales with ground VLBI
\citep{kr90}. \cite{lo98} observed the source at 5~GHz with VSOP
and also reported oscillations of the ridge line. Identifying
these structures with Kelvin-Helmholtz (KH) modes, they were able
to derive the basic parameters of the flow. High dynamic range
VSOP and VLBA (Very Long Baseline Array of National Radio
Astronomy Observatory, USA) observations of 0836+710 at
$1.6\,\rm{GHz}$ show the presence of an oscillation at a
wavelength as long as $\sim100\,\rm{mas}$ \citep{lo06}, that
cannot be readily reconciled with the jet parameters given in
\cite{lo98}. \cite{pl07} have shown that the presence of a shear
layer allows to fit all the observed wavelengths within a single
set of parameters, assuming that they are produced due to KH
instabilities growing in a cylindrical outflow. In this picture,
the longest mode corresponds to a surface mode growing in the
outer layers, whereas the shorter wavelengths are identified with
body modes developing in the inner radii of the jet.

We report here on further progress of this investigation. We use
different observations of the radio jet in the quasar S5\,0836+710
with different VLBI networks (VSOP, global VLBI, VLBA and EVN) at several
frequencies (1.6, 2, 5, 8, 15, 22 and 43~GHz) in order to obtain
the ridge lines of the jet and compare their evolution in time at
the different scales given by each frequency.

In Section~\ref{ridge} we present some of the ridge lines obtained from the
different epochs and frequencies and explain the way in which the
fits will be carried out. In this section we also show how this
kind of functions can be used as a predictive tool for
observations of sub-parsec scale jets. Finally, we present our
conclusions in Section~\ref{sum}.

\section{Ridge lines}\label{ridge}

The use of observations of the jet at different wavelengths and
epochs is critical to our study. The different wavelengths can
help to derive information on the different structures arising in
different radial regions of the jet \citep{pl07} and having
different epochs can be useful for determining the motions of
these ridge lines. We have thus used data from VLBA and VSOP at 
1.6 and 5~GHz \citep{lo06} at two different epochs, one epoch 
at 1.6~GHz from EVN (Perucho et al. in preparation), one epoch of 
simulataneous global VLBI (including VLBA) observations at 2 and 8~GHz 
(01/1997, Pushkarev \& Kovalev, in preparation), and two epochs from VLBA at 
8, 22 and 43~GHz (Perucho et al., in preparation). Finally, we have made 
use of 11 epochs between 1998 and 2007 from the 2\,cm~VLBA/MOJAVE database at 15~GHz. 


At every observing frequency, different structure wavelengths are
detected in the jet. When the different epochs are plotted
together, a displacement of the ridge line is observed at all
frequencies. Whether this displacement could be attached to
experimental errors was tested by using a straight jet from the
MOJAVE database. The results obtained showed that the
differences observed between epochs in 0836+710 were much larger
than the ones seen in the straight jet, which make us confident on the
displacements being due to real or apparent motions.

Assuming thus that the observed differences between ridge line
positions in time, for a given frequency, are due to motions of structures in the jet, and that these structures are due to pressure enhancements produced by
any type of instability, we can try to fit them to a helix and
derive its properties. The equations that determine the shape of
helix in three dimensions, depending on the $z$ coordinate and
time are $x(z,t) = A(z)\,\,
cos(2\pi\,(z+v_w\,t)/\lambda\,+\,\phi)$ and $y(z,t) = A(z)\,\,
sin(2\pi\,(z+v_w\,t)/\lambda\,+\,\phi)$, where $v_w$ is the wave
speed, $\lambda$ is the wavelength, $\phi$ is the initial phase
and $A(z)$ is the amplitude that depends on the growth rate of the
instability $l_i$. The amplitude is fixed to 0 at origin, as the
position of the wave does not change at the core: $A(z) =
A_0\,exp(z/l_i)\,sin(2\pi\,z/\lambda)$. The expressions given
above for the $x$ and $y$ coordinates have to be corrected for
projection and relativistic effects and rotated in order to align
it with the direction of the observed jet. This depends on the
viewing angle to the jet and the position angle of the jet, which
are known, but also on the wave speed.

The fits depend on five parameters: $A_0$, $l_i$, $\lambda$, $v_w$
and $\phi$, that will determine the properties of the wave and can
thus help in obtaining information about the jet flow itself, via
linear stability analysis of KH or current driven instabilities.
The different epochs are taken into account via the time
parameter, so all epochs are to be fitted at a time to a single
moving structure.

At 43~GHz, we observe a lateral displacement of the ridge line,
similar to that reported by \cite{ag07} and references therein for
the jet in NRAO~150. Using the equations given in the previous
section as a model for a short arbitrary mode with reasonable
properties of a short body mode, we can produce an artificial
helix and use the time dependence to check its apparent motion for
an observer. The results, displayed in Fig.~1, show that the
regular wave motion of a pressure enhancement in a jet can result
in an apparent transversal motion of the jet itself if the
observed region corresponds only to this pressure maximum. This
latter statement has to be considered in the frame of the high
frequency at which these motions are observed, which may result in
only a small part of the jet---that with the highest energies---being 
seen. The superluminal nature of this transversal motions remains to be explained.

\begin{figure}[!h] \label{fig2}
\begin{center}
\includegraphics[width=0.43\textwidth,angle=0,clip=true]{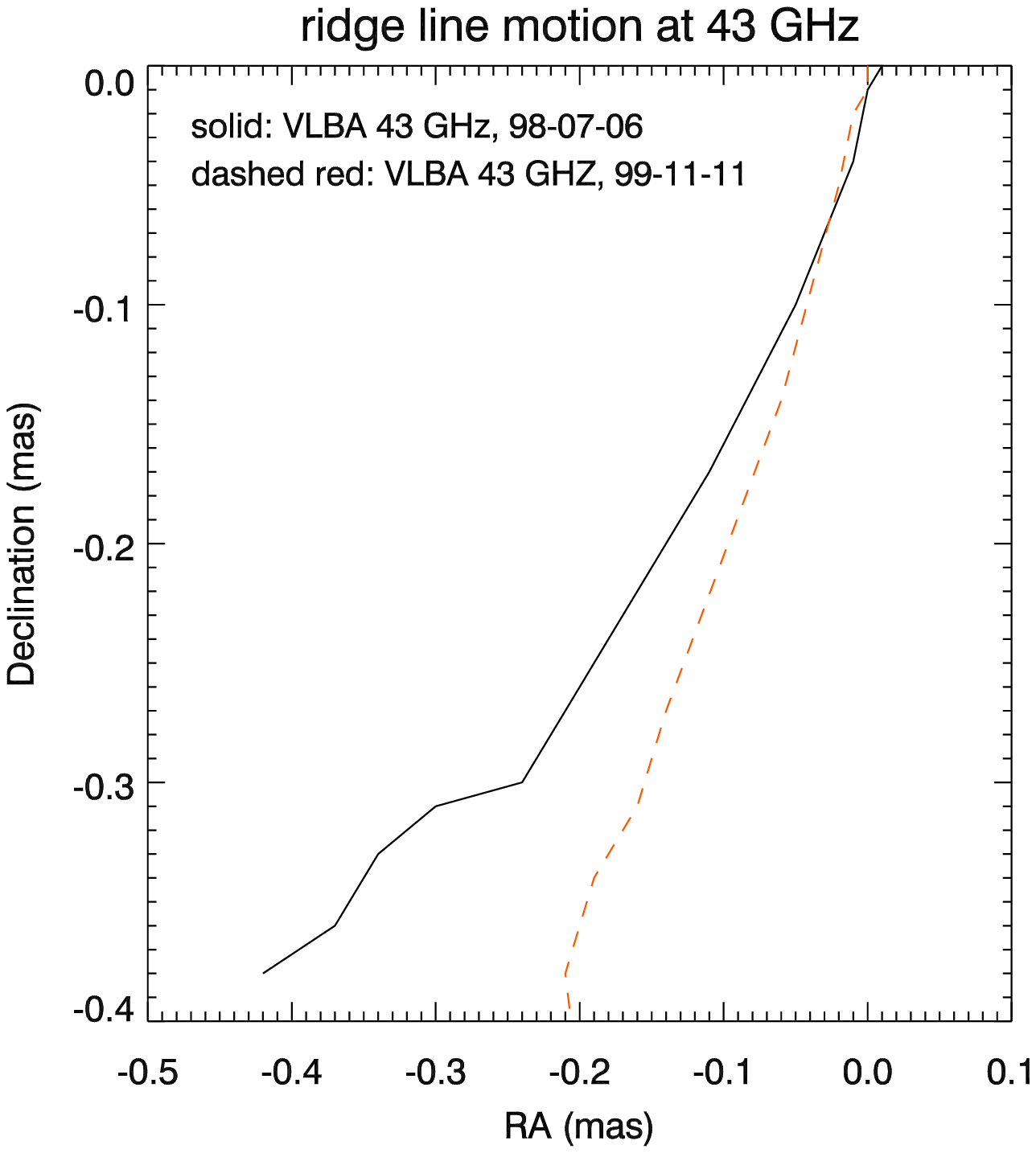}\\
\includegraphics[width=0.43\textwidth,angle=0,clip=true]{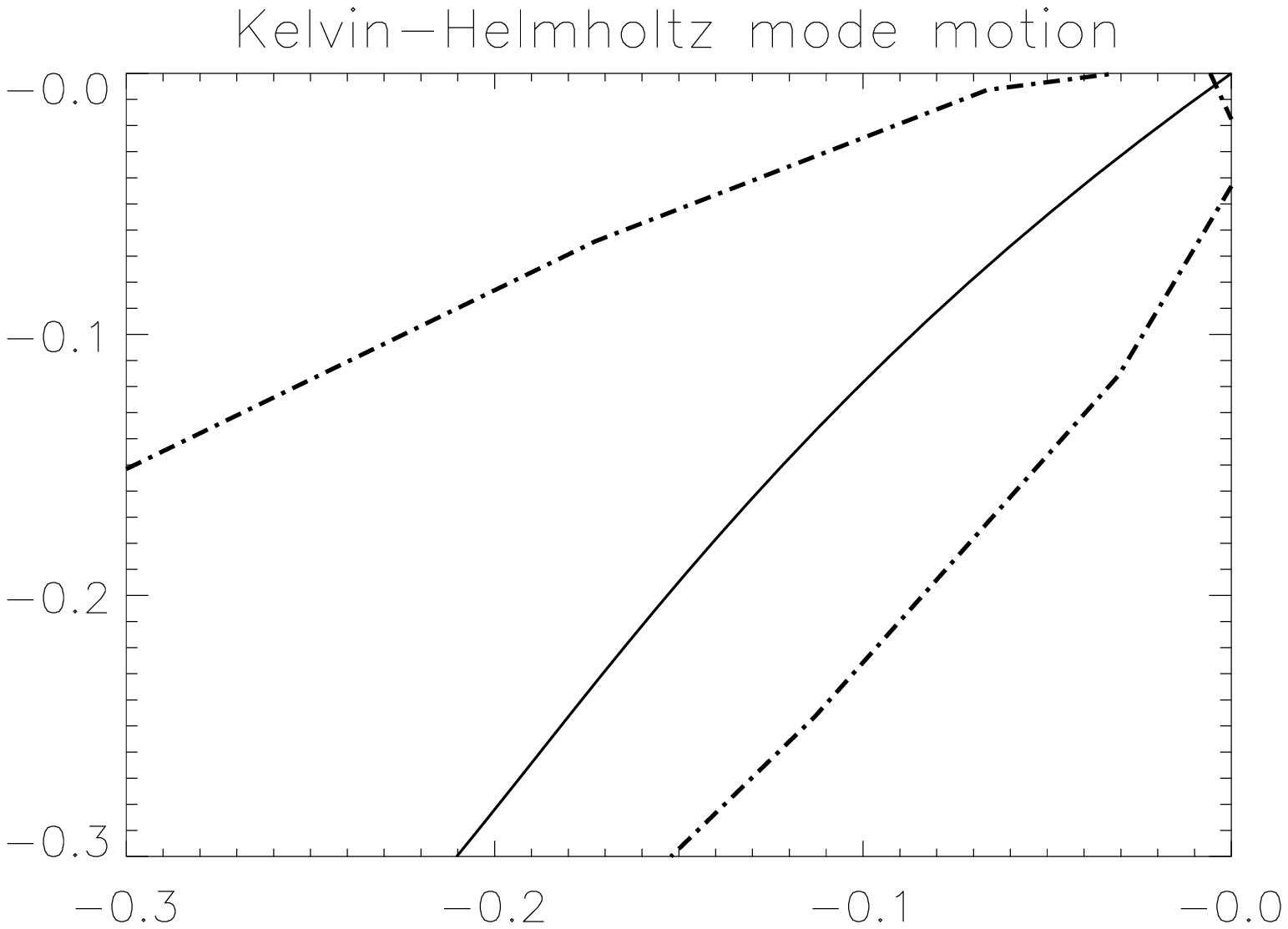}
\includegraphics[width=0.43\textwidth,angle=0,clip=true]{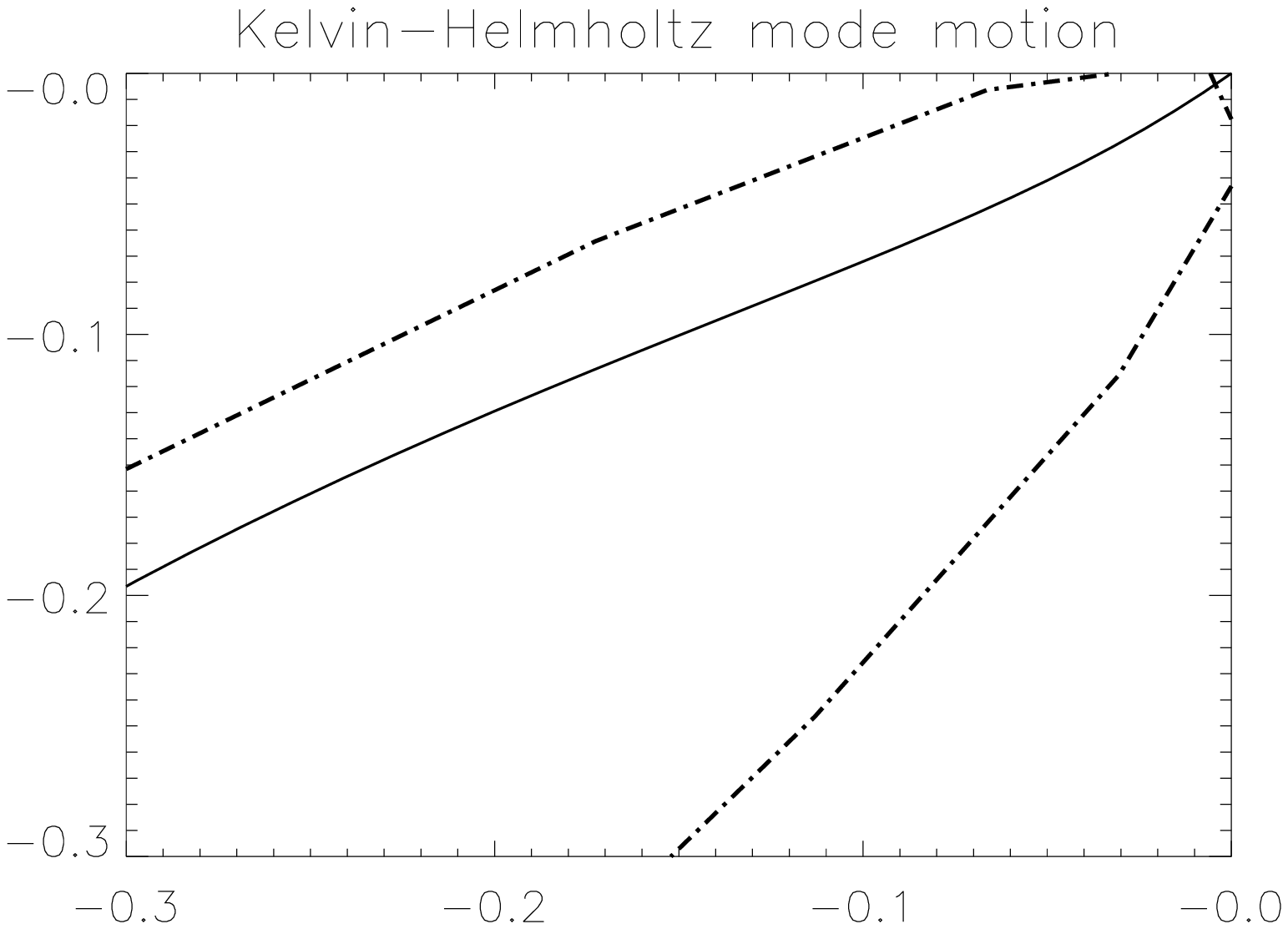}
\end{center}
\caption{The top panel shows the ridge lines of the 43~GHz images
in epoch 07/1998 (solid) and 11/1999 (dashed). The two bottom
panels show the apparent motion of a short-wavelength structure
ridge line seen at the same scales as the observations at this
frequency. The dash-dot lines indicate possible real boundaries of
the flow.}
\end{figure}

\section{Summary}\label{sum}
We are able to explain the structures observed in the jet at
several frequencies assuming that a shear layer exists between the
jet and the ambient medium. The observed structure of jets changes
with frequency, but also do the hydrodynamics \citep{pe06,pl07}.
The fits to the ridge line at different epochs and frequencies can
give insight on the basic properties of the instabilities,
identify modes and provide better estimates of the physical
parameters of jets. At the same time this can help testing the KH
and current driven instability scenarios for the parsec and
kiloparsec scales, depending on the studied frequencies. We
observe helical structures at all frequencies in the jet in
0836+710. We claim that they cannot all be due to precession of
the same object. The periods for these waves range from several years to
$10^7$~years. The origin of these motions has to be addresses in
the future and on the light of the results of this work. The
rotation of a structure, such as a wave pattern produced by an
instability can generate the observed radial motion of jets at the
highest frequencies. Superluminal velocities remain to be
explained (from observations or theory). High resolution
observations like those planned in the VSOP-2 project will be critical 
in this effort.

\acknowledgements
This work was supported in part by the Spanish
Direcci\'on General de Ense\~nanza Superior under grants
AYA-2001-3490-C02 and AYA2004-08067-C03-01. M.P. is supported by a
postdoctoral fellowship of the Generalitat Valenciana ({\it Beca
Postdoctoral d'Excel$\cdot$l\`encia}). Y.K. is a research fellow of the
Alexander von Humboldt Foundation. This research has made use of the data from the MOJAVE \citep{lh05} and 2cm Survey \citep{k04} programs.

\end{document}